# Gradient induced liquid motion on laser structured black Si surfaces

I. Paradisanos[1,2], C. Fotakis[1,2], S. H. Anastasiadis[1,3], E. Stratakis[1,4,a]

[1] *Institute of Electronic Structure and Laser (IESL), Foundation for Research and Technology-Hellas (FORTH), Heraklion, 71003, Greece*

[2] *Physics Department, University of Crete, Heraklion, 71003, Greece*

[3] *Chemistry Department, University of Crete, Heraklion, 71003, Greece*

[4] *Materials Science and Technology Department, University of Crete, Heraklion, 71003, Greece*

This letter reports on the femtosecond laser fabrication of gradient-wettability micro/nano-patterns on Si surfaces. The dynamics of directional droplet spreading on the surface tension gradients developed is systematically investigated and discussed. It is shown that microdroplets on the patterned surfaces spread at a maximum speed of 505 mm/sec, that is the highest velocity demonstrated so far for liquid spreading on a surface tension gradient in ambient conditions. The application of the proposed laser patterning technique for the precise fabrication of surface tension gradients for open microfluidic systems, liquid management in fuel cells and drug delivery is envisaged.

a) Corresponding author:stratak@iesl.forth.gr

Si is the most widely used material for semiconductor devices, especially in integrated circuits used in microelectronics and computers. One less obvious but interesting application of Si is that it is commonly used for microfluidic[1,2] and electronics cooling[3] applications. In such applications, the capability of modifying the surface wetting properties of Si is incontrovertibly important.[4] Recently, we have demonstrated that femtosecond (fs) laser structured, black Si (BS) surfaces exhibit, among other interesting properties, remarkable wetting characteristics.[5] In particular, we have shown that it is possible to preferentially tune the wettability of Si from superhydrophobic[6] to superhydrophilic[7] through a proper combination of surface topography and chemistry. Furthermore, depending on the type of functional coating deposited onto BS surfaces, pH-[8], photo-[9], and electro-[10] responsiveness was attained, which can potentially be important for microfluidic applications.

Besides this, currently there has been great research interest to induce liquid motion onto solid surfaces via using chemical[11-14], thermal[15-18], photochemical[19-20] and electrochemical[21] methods. Most of the techniques developed are based on surface tension or wettability gradients, i.e. surfaces for which the solid-liquid surface energy, $\gamma_{SL}$, gradually changes towards a specific direction. As a result of the gradient surface tension and the subsequent differences in dynamic contact angles (CAs), liquid motion towards the direction of decreasing $\gamma_{SL}$ is induced. Apart from the various types of chemical modifications, a surface energy gradient can be also achieved by a well-defined roughness gradient.[22-24] Typical liquid velocities attained using the different approaches reported to date, are ranging from micrometers to a few millimeters per second, which are too low for practical applications, including liquid thermal management in fuel cells and microfluidic devices.[17] In this respect, the development of a liquid propulsive technique enabling passive transport of small and/or big volumes of fluids at high velocities is desirable.

In this work we demonstrate a novel methodology to fabricate a well-defined gradient wettability pattern on Si, exhibiting the highest velocity reported so far for liquid spreading on a surface tension gradient in ambient conditions. The surfaces developed comprise fs laser patterned BS areas consisting of quasi-periodical arrays of microcones (MC).[25] By gradually decreasing the irradiation fluence, the MC characteristics and thus the surface topology is affected in a way that the initially textured hydrophobic regions gradually turn to more hydrophilic and finally to superhydrophilic. Accordingly, the driving mechanism for liquid motion is provided by the gradual alteration of the surface roughness at the micro- and the nano- scale, giving rise to respective wettability changes. Our technique is simple and contrary to previously reported methods it does not require deposition of additional coatings. More importantly, it enables high-velocity spontaneous motion of water volumes against gravity, i.e. even when the surface is vertically positioned and the droplet is running uphill. Our results demonstrate the potential application of laser patterning for the fabrication of wettability gradients in open microfluidic systems.

Single crystal n-type (phosphorus doped) silicon wafers (100) with a resistivity of $\rho$=2-8$\Omega$cm were mounted into a vacuum chamber evacuated down to a residual pressure of $5\times10^{-2}$ mbar by means of a rotary pump. A constant $SF_6$ pressure of 500 Torr was maintained during the laser patterning process, via a precision micro valve system. The laser source used was a regenerative Ti:Sapphire ($\lambda$ = 800nm) system delivering 200 fs pulses at a repetition rate of 1kHz. The chamber was positioned on a high-precision, computer controlled, X-Y translation stage, normal to the incident beam. A motorized neutral density filter wheel was synchronized to the translation stages, allowing accurate control of the fluence and number of pulses irradiating a specific sample area. Using this setup, the wettability gradient (Figure 1a) can be obtained by surface treatment using a starting incident fluence of 1.31 J/cm$^2$ which is gradually decreased to 0.25 J/cm$^2$ (Figure 1(b)) with a step of 0.02 J/cm$^2$, while maintaining a constant average number of pulses (500) per irradiation spot. In our case, the dimensions of the gradient textured area were 3000μm by 2000 μm. Subsequently, the fluence was set to a value of 1.72 J/cm$^2$ and an area of the same width but with higher length (3000μm by 4200μm) was patterned using this constant fluence. Following the irradiation process, the sample was thermally oxidized at 1000$^o$C for 30 min and subsequently, only the gradient surface region was treated by a 10% HF aqueous solution for 10 minutes. As a result, the HF-treated gradient textured area exhibited gradient hydrophobicity, while that irradiated at constant fluence and subsequently oxidized behaved as superhydrophilic.

For the fabrication of the simple microfluidic element depicted in Figure 4a, the two gradient patterns (3000μm by 6000μm each) were fabricated by gradually changing the laser fluence from 1.72 J/cm$^2$ to 0.30 J/cm$^2$. The patterns were separated by an area of 8000μm by 8000μm fabricated at a constant laser fluence of 1.72 J/cm$^2$. The sample was subsequently placed in the oxidation oven at 1000$^o$C for 30 minutes without a final HF treatment.

The morphology of the different textured Si areas was characterized by scanning electron microscopy (SEM). An image-processing algorithm was implemented in order to obtain quantitative information concerning the macroscopic characteristics of the structures formed, i.e. MC density, height and distribution, from top and cross-sectional SEM pictures of the structured areas. The experimental investigation of the liquid motion on the fabricated surfaces was performed by placing a 4μl distilled water droplet onto the sample and subsequent monitoring of the resulting droplet motion by a high speed camera (Casio Exilim Pro EX-F1) at 600 frames per second (fps). The droplet's displacement has been calculated via analysis of the video frames monitoring the evolution of the droplet's position over time. The respective velocity was calculated from the time derivative of the droplet's volume displacement. Finally, the static contact angles on extended areas structured at different fluences were measured by a tensionmeter (Dataphysics OCA-40 CA/ST meter).

Micropatterning of the substrates used in this study was performed by fs laser structuring of crystalline Si wafers. This technique offers the advantage of patterning Si surfaces with periodic arrays of MCs, while offering high accuracy and reproducibility.[7] Upon increasing the laser fluence, the height of the MCs along with the number of nanoscale protrusions decorating the cones increases, resulting in a significant enhancement of the overall roughness. At the same time the initially hydrophilic Si surface becomes hydrophobic, while hydrophobicity is further enhanced upon increasing laser fluence. Figure 1(a) shows an optical microscopy image of the laser structured gradient wettability pattern on Si fabricated to induce liquid motion. The pattern is consisted of two different but connected areas: The gradient area, treated with gradually decreasing laser fluence (upper side), which is optically distinguished due to the differences of the MCs height and density, giving rise to light absorption variations. Specifically, as the hydrophobicity is decreased the surface becomes brighter and gradually turns from black into grey. In the lower side, the superhydrophilic area can be seen which was uniformly structured at constant laser fluence and subsequently was thermally oxidized. Schematic representations of the meander motif laser scan as well as the fluence variation and the dimensions of the two differently treated areas are presented in Figure 1(b). The corresponding SEM images of a part of the gradient area as well as the onset of the oxidized area, are depicted in Figure 1(c). Note that the MCs located into the superhydrophilic area are brighter in view than the non-oxidized ones, due to the oxide formation on the surface. It is clear that the continuous variation of the laser fluence is affecting the height and density of the MCs. This is well demonstrated in Figure 1(d), showing the linear decrease of the MCs' height, accompanied by a corresponding increase of their density as a function of the position along the gradient. The spatial variation of the geometrical characteristics of the MCs gives rise to a corresponding CA variation in a way that the liquid volume is forced to spread from the hydrophobic to the less hydrophobic part of the gradient and finally to the superhydrophilic part. The respective static CA measured on extended areas structured at different fluences are also shown in Figure 2(d).

Figure 2 presents a selected time sequence of snapshots of a 4µl droplet spreading along a horizontal, a 30$^o$ and a 90$^o$ tilted substrate as a function of time. It is observed that the droplet spreads with high velocity towards the more wettable part of the gradient, while the dynamic contact angle is continuously decreasing. As soon as the droplet enters the superhydrophilic area it acquires a remarkable maximum spreading velocity of 505 mm/s. This is the highest velocity reported so far for a droplet spreading on a surface tension gradient at room temperature.[26] Such high velocity can be attributed to the special design of the gradient, particularly to the initially gradual change of the CA from 133 to 93 degrees, followed by the steep fall to 0 degrees. The corresponding evolutions of the droplet position and velocity along the gradient are displayed in Figure 3(a) and 3(b) respectively. Experiments performed on inclined surfaces showed that gravity affects droplet spreading and as shown in Figure 3(b) the maximum attainable velocity decreases to 450 mm/sec and 202 mm/sec for 30$^o$ and 90$^o$ substrate inclination angles respectively. These velocity values for water volumes spreading uphill onto a surface are also surprisingly high, compared with the respective values attained in the literature.[11-21]

As a proof of the potential application of the laser fabricated wettability gradients in open microfluidic elements, we have patterned onto a Si wafer a double gradient that can be used to drive a liquid volume supplied to a supehydrophilic area towards a specified direction. As shown in Figure 4a, the element fabricated comprises two gradient arms, indicated as regions '1' and '3', separated by a superhydrophilic region, denoted as '2', which facilitates as a water reservoir. Specifically, the region '1'behaves as a hydrophilic gradient that starts from hydrophilic and ends to a superhydrophilic region (at the side of the reservoir), whereas the region '3' starts from a superhydrophilic and ends to a hydrophilic (at the side of the reservoir) area.

Figure 4b presents a selected series of snapshots of a water droplet supplied to area '2', showing that the liquid is preferentially moving towards the more hydrophilic gradient arm. In particular, the droplet spreads from a superhydrophilic area (0 degrees) to a gradient of increasing hydrophobicity; at the same time it is hindered to spread to the opposite direction by a gradient of decreasing hydrophobicity. Upon progressive supply of the reservoir with water, the liquid volume is directed only to this specific arm at a velocity of 27 mm/sec. It should be emphasized that the observed functionality was the main aim for the fabrication of this element, while the gradient properties of this system were not designed to obtain a maximum droplet velocity.

The experimental results obtained can be interpreted in the framework of a simplified model accounting for the equilibrium of the different types of forces governing the droplet motion onto a surface tension gradient, describes by Equation 1.

$$\sum F = F_{\text{act}} - F_{\text{hys}} - F_{\text{vis}} = ma \qquad (1)$$

The actuation force, $F_{\text{act}}$, that moves the droplet onto the surface originates from the variable surface energy and thus wettability of the liquid-solid interface, whereas the resistance force comprises two components: one is the hysteresis force, $F_{\text{hys}}$, resulting from the hysteresis phenomenon taking place before a droplet begins to move, and the other is the viscous force, $F_{\text{vis}}$, of the liquid, taking place during the droplet motion.

Specifically, the actuation force is given by the gradient of surface energy along the direction of motion and can be expressed by[27]

$$F_{\text{act}} = -\frac{d\Delta G}{dx} \cong \pi R_{\text{b}}^2 \gamma_{\text{LV}} \left(\frac{d \cos\theta}{dx}\right) \qquad (2)$$

, where $\gamma_{LV}$ represents the surface tension of a droplet, $R_b$ is the droplet's base radius in contact with the solid, and $\theta$ is the position-dependent CA of the liquid droplet on the solid surface, which decreases in the same direction that the droplet moves. $R_b$ was precisely measured via top and side view snapshots of droplet's profile, obtained during the respective CA measurements. Contact angle hysteresis represents a moving barrier that the droplet should overcome prior its motion. The corresponding hysteresis force can be expressed as:[28, 29]

$$F_{hys} = \gamma_{LV}(\cos\theta_R - \cos\theta_A)w_{c,eff} \qquad (3)$$

, in which $\cos\theta_A$ and $\cos\theta_R$ are the dynamic advancing and receding contact angles at the instant at which a droplet begins to move, respectively, and $w_{c,eff}$ is the effective contact length for a droplet placed on the textured surface. As the droplet moves on the gradient-textured area, the resistance force resulting from the viscous force can be expressed as[29, 30]

$$F_{vis} \cong 3\pi\eta R_b V \ln\left(\frac{X_{max}}{X_{min}}\right) \qquad (4)$$

, where $\eta$ is the viscosity of the liquid and $V$ is the velocity of the moving droplet. $X_{max}$ and $X_{min}$ are the lengths of a liquid: generally, $X_{min}$ is the length of a molecule in liquid phase and $X_{max}$ represents the radius of the droplet. Based on Equation 1, when the droplet is moving steadily, the driving force acting on the droplet is equal to the viscous force (i.e. $F_{act} = F_{vis}$) and the velocity of the moving droplet can be expressed as[25, 27]

$$V \cong \frac{\gamma_{LV} R_b}{3\eta \ln(X_{max}/X_{min})}\left(\frac{d\cos\theta}{dx}\right) \qquad (5)$$

By transforming equation (5) we end up in the following expression

$$V^2 \cong \frac{\gamma_{LV} R_b}{3\eta \ln(X_{max}/X_{min})}\left(\frac{d\cos\theta}{dt}\right) \qquad (6)$$

The factor $\frac{d\cos\theta}{dt}$ can be calculated from the respective video frames of the water droplet while it is moving onto the gradient texture. Therefore, Equation (6) can be used to calculate the evolution of the droplet velocity over time and directly compare it with that determined from the experiment. The velocity values calculated from Equation (6) for the $0^o$-tilt case, shown in Figure 3(b), are in good agreement with the ones determined experimentally. It should be noted that the above formulation is applicable while the liquid volume is within the gradient area and only in the case of $0^o$ tilt, where the gravitational force is not participating to the forces that govern the droplet motion. Accordingly, the calculated velocity values shown in Figure 3(b) correspond to droplet motion in that case. From the results shown in Figure 3(b), it can be concluded that the simplified model we have used can pretty well interpret our experimental observations.

In summary, we have developed a novel technique to fabricate gradient wettability patterns on Si that exhibit remarkable liquid spreading velocities, even in the case of vertically positioned surfaces. The gradual alteration of the surface roughness in the micrometer and nanometer scale gives rise to a gradient surface tension that provides the driving mechanism for liquid motion. Our experimental observations were well interpreted by a simplified model that describes the equilibrium of the different forces governing the droplet motion on a surface tension gradient. The propulsive approach proposed here, can be a promising option in open microfluidic systems, liquid management in fuel cells and drug delivery applications.[31]

**ACKNOWLEDGEMENTS**

This work was performed in the framework of the PROENYL research project, Action KRIPIS, project No MIS-448305 (2013SE01380034) that was funded by the General Secretariat for Research and Technology, Ministry of Education, Greece and the European

**Figures**

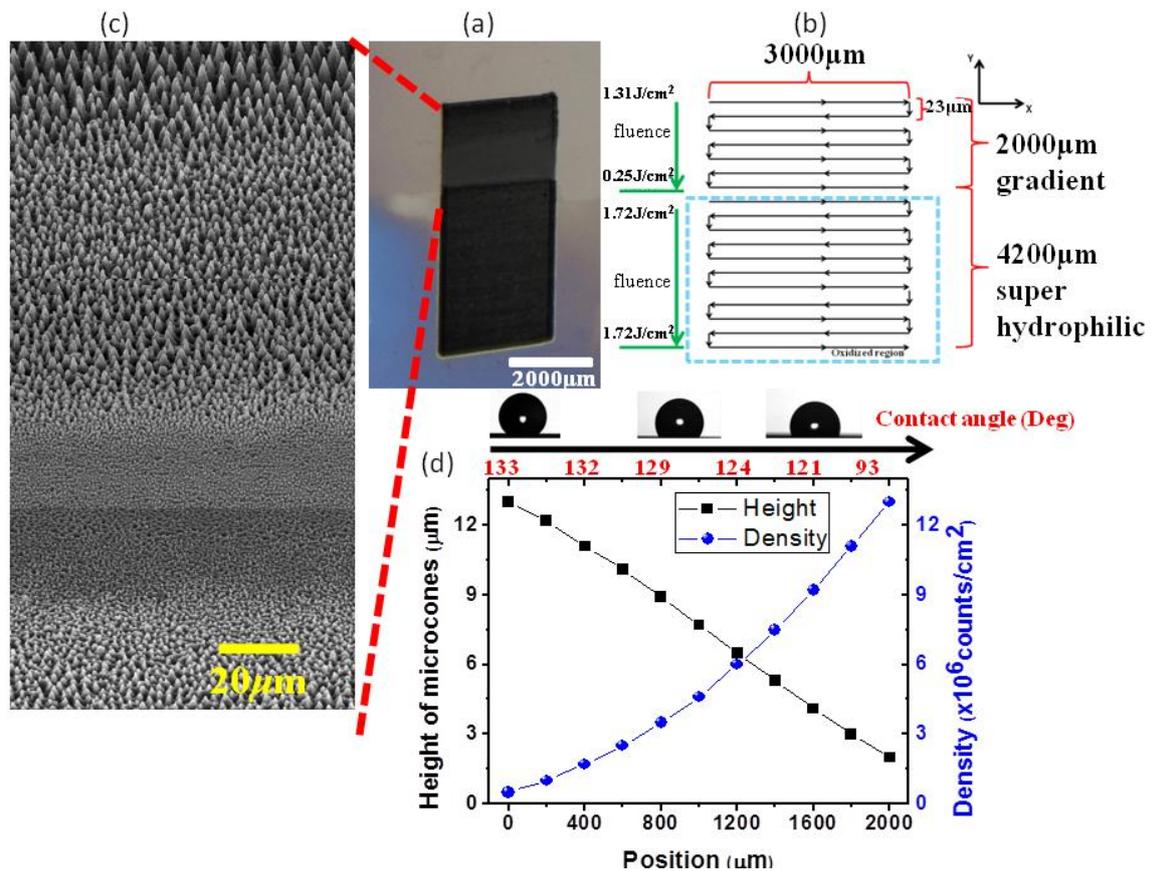

**Figure 1:** a) Optical image of the structured Si substrate. The gradient and super-hydrophilic part are visible; b) Schematic representation of the laser patterning process, the laser fluence variation and the dimensions of the textured areas; c) Scanning Electron Microscopy image (SEM) showing the variation of MCs along the laser-patterned surface gradient; note the brighter shade of the oxidized MCs at the bottom; d) MCs height and density as a function of the position along the gradient area. The static CA values and typical droplet profiles are shown on the upper side.

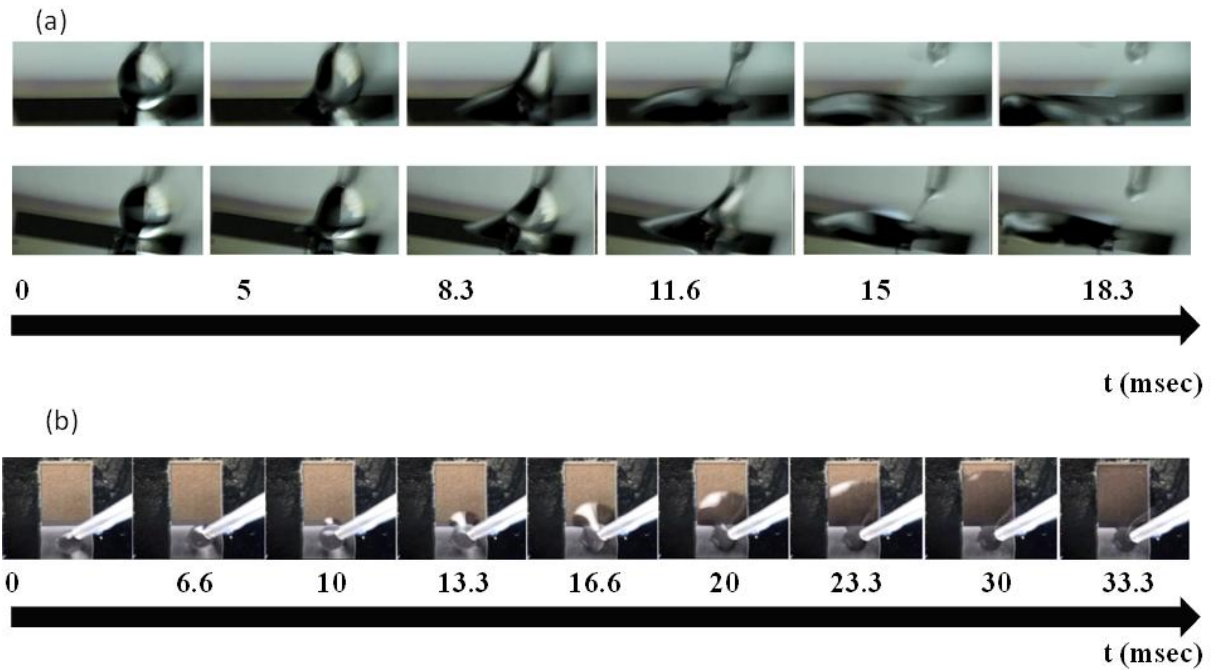

**Figure 2:** (a) Selective snapshots taken from high-speed camera videos, of a 4μl water droplet spreading on the laser structured Si surface for inclination angles of 0º (up) and 30º (down); (b) 4μl water droplet running uphill on the vertically positioned structured substrate.

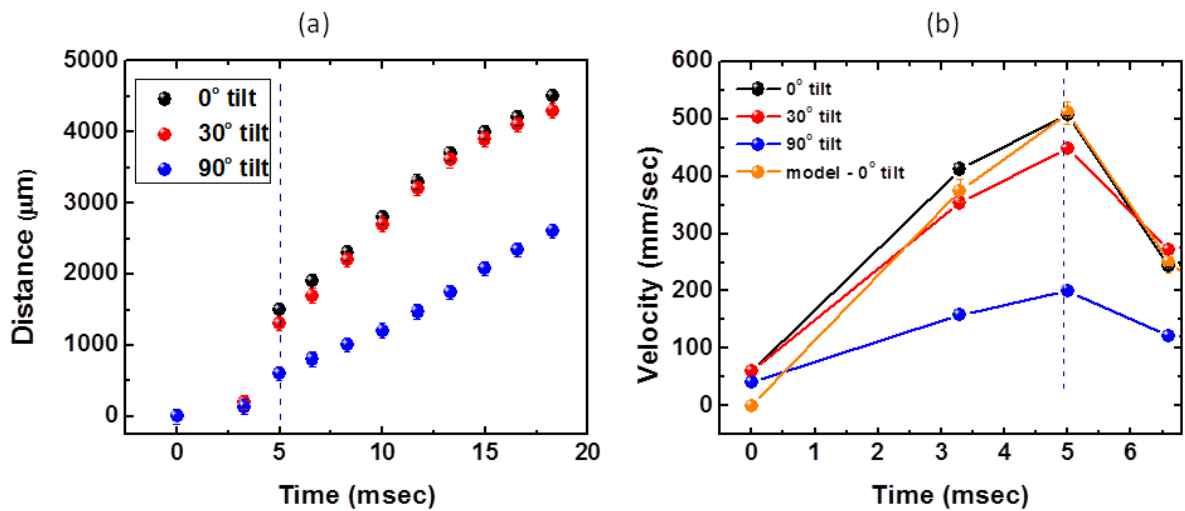

**Figure 3:** (a) Evolution of a 4μl water droplet position as a function of time for three different substrate inclination angles. (b) Evolution of droplet's velocity for the three different substrate inclination angles; the corresponding velocity values calculated for the 0° inclination angle case using a simplified model described by Equation (6) is also presented for comparison. The dashed lines correspond to the time, at which the droplet acquires maximum velocity.

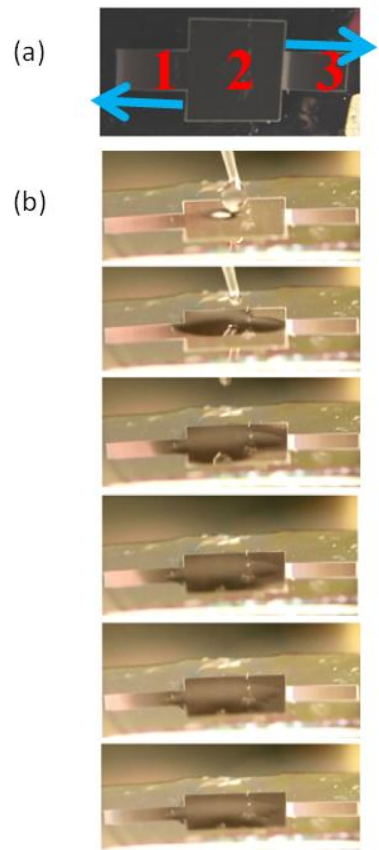

**Figure 4:** a) Optical image of the double gradient element structured onto a Si substrate. Three different regions are distinguishable: in region 1 the arrow denotes the gradient decrease of the hydrophilicity whereas in region 3 the arrow corresponds to the gradient

increase of the hydrophilicity. Region 2 is a superhydrophilic reservoir that receives the water volume; (b) Snapshots of a 4μl water droplet spreading on the element described in a).

## References

[1] P. Gravesen, J. Branebjerg, O.S. Jensen, *J. Micromech. Microeng.* **3**, 168-182(1993).

[2] G.M. Whitesides, *Nature* **442**, 368–373 (2006).

[3] D. Erickson and D. Li, *Anal. Chim.* **507**, 11–26 (2004).

[4] R.N.J. Wenzel, *Phys. Colloid Chem.* **53**, 1466–1467 (1949).

[5] E. Stratakis, A. Ranella and C. Fotakis, *Biomicrofluidics* **5**, 013411 (2011).

[6] V. Zorba, E. Stratakis, M. Barberoglou, E. Spanakis, P. Tzanetakis, S. H. Anastasiadis and C. Fotakis, *Adv. Mater.* **20**, 4049-4054 (2008).

[7] E. Stratakis, *Science of Adv. Mater.* **4**, 407-431 (2012).

[8] E. Stratakis, Anca Mateescu, M. Barberoglou, M. Vamvakaki, C. Fotakis and S. H. Anastasiadis, *Chem. Commun.* **46**, 4136-4138 (2010).

[9] E. L. Papadopoulou, M. Barberoglou, V. Zorba, A. Manousaki, A. Pagkozidis, E. Stratakis and C. Fotakis, *J. Phys. Chem. C* **113**, 2891-2895 (2009).

[10] M. Barberoglou, V. Zorba, A. Ragozidis, C.Fotakis and E. Stratakis, *Langmuir* **26**, 13007-13014 (2010).

[11] H. A. Stone, A. D. Stroock and A. Adjari, *Annu. Rev. FluidMech* **36**, 381 (2004).

[12] Y.Sumino, N. Magome, T. Hamada and K. Yoshikawa, *Phys. Rev. Lett.* **94**, 068301 (2005).

[13] Y. Zheng, H. Bai, Z. Huang, X. Tian, F. Nie, Y. Zhao, J. Zhai and L. Jiang, *Nature (London)* **463**, 640 (2010).

[14] O. Bliznyuk, P. H. Jansen, E. S. Kooij, H. J. W. Zandvliet and B. Poelsema, *Langmuir* **27**, 11238–11245 (2011).

[15] A. M. Cazabat, F. Heslot, S. M. Troian and P. Carles, *Nature (London)* **346**, 824 (1990).

[16] M. J. Hancock, J. He, J. F. Mano and A. Khademhosseini, *Small* **7**, 892 (2011).

[17] S. Daniel, M. K. Chaudhury and J. C. Chen, *Science* **291**, 633 (2001).

[18] P. Lazar and H. Riegler, *Phys. Rev. Lett.* **95**, 136103 (2005).

[19] K. Ichimura, S. Oh and M. Nakagawa, *Science* **288**, 1624 (2000).

[20] S. Abbott, J. Ralston, G. Reynolds and R. Hayes, *Langmuir* **15**, 8923 (1999).

[21] B. S. Gallardo, V. K. Gupta, F. D. Eagerton, L. I. Jong, V. S. Craig, R. R. Shah and N. L. Abbott, *Science* **283**, 57 (1999).

[22] M. Prakash, D. Quéré and J. W. M. Bush, *Science* **320**, 931 (2008).

[23] N. A. Malvadkar, M. J. Hancock, K. Sekeroglu, W. J. Dressick and M. C. Demirel, *Nat. Mater.* **9**, 1023 (2010).

[24] M. M. Weislogel, J. A. Baker and R. M. J. Jenson, *Fluid Mech.* **685**, 271 (2011).

[25] V. Zorba, L. Persano, D. Pisignano, A. Athanassiou, E. Stratakis, R. Cingolani, P. Tzanetakis and C. Fotakis, *Nanotechnology* **17**, 3234 (2006).

[26] H. S. Khoo and F. G. Tseng, *Appl. Phys. Let.* **95**, 063108 (2009).

[27] S. Daniel and M. K. Chaudry, *Langmuir* **18**, 3404-3407 (2002).

[28] G. J. Furnidge, *Colloid Interface Sci.* **17**, 309-324 (1962).

[29] J. T. Yang, Z. H. Yang, C. Y. Chen and D. J. Yao, *Langmuir* **24**, 9889-9897 (2008).

[30] H. Suda and S. Yamada, *Langmuir* **19**, 529-531 (2003).

[31] N. T Nguyen et al., Advanced Drug Delivery Reviews **65**, 1403 (2013).